\documentclass[aps,twocolumn,amsmath,amssymb,nofootinbib,superscriptaddress,floatfix,longbibliography]{revtex4-1}
\usepackage[dvips]{graphicx}
\usepackage{latexsym}
\usepackage{amsmath}
\usepackage{amsfonts}
\usepackage{amssymb}
\usepackage{bm}
\usepackage{color}
\usepackage{txfonts}
\usepackage{float}
\usepackage{url}
\usepackage[colorlinks=true, urlcolor=blue, linkcolor=blue, citecolor=blue, pdftex]{hyperref}
\usepackage{ulem}
\usepackage{physics}
\usepackage{tikz}

\begin{document}
	\newcommand{\fig}[2]{\includegraphics[width=#1]{#2}}
	\newcommand{\pprl}{Phys. Rev. Lett. \ }
	\newcommand{\pprb}{Phys. Rev. {B}}

\title{Spin-Lattice Relaxation in Two-Dimensional Superconducting BKT Transition}

\author{Wei-Wei Yang}
\affiliation{Beijing National Laboratory for Condensed Matter Physics and Institute of Physics,
	Chinese Academy of Sciences, Beijing 100190, China}

\author{Shao-Hang Shi}
\affiliation{Beijing National Laboratory for Condensed Matter Physics and Institute of Physics,
	Chinese Academy of Sciences, Beijing 100190, China}
\affiliation{School of Physical Sciences, University of Chinese Academy of Sciences, Beijing 100190, China}

\author{Zongsheng Zhou}
\affiliation{Beijing National Laboratory for Condensed Matter Physics and Institute of Physics,
	Chinese Academy of Sciences, Beijing 100190, China}

\author{Zi-Xiang Li}
\email{zixiangli@iphy.ac.cn}
\affiliation{Beijing National Laboratory for Condensed Matter Physics and Institute of Physics,
	Chinese Academy of Sciences, Beijing 100190, China}
\affiliation{School of Physical Sciences, University of Chinese Academy of Sciences, Beijing 100190, China}
    
\author{Kun Jiang}
\email{jiangkun@iphy.ac.cn}
\affiliation{Beijing National Laboratory for Condensed Matter Physics and Institute of Physics,
	Chinese Academy of Sciences, Beijing 100190, China}
\affiliation{School of Physical Sciences, University of Chinese Academy of Sciences, Beijing 100190, China}

\author{Jiangping Hu}
\email{jphu@iphy.ac.cn}
\affiliation{Beijing National Laboratory for Condensed Matter Physics and Institute of Physics,
	Chinese Academy of Sciences, Beijing 100190, China}
\affiliation{Kavli Institute of Theoretical Sciences, University of Chinese Academy of Sciences,
	Beijing, 100190, China}
 \affiliation{New Cornerstone Science Laboratory, 
	Beijing, 100190, China}
    
\date{\today}
\begin{abstract}

Two-dimensional superconductors undergo a Berezinskii–Kosterlitz–Thouless transition driven by vortex–antivortex unbinding, yet experimental signatures beyond transport remain limited. Here, we show that the spin-lattice relaxation rate provides a direct probe of this transition. In a 2-dimensional $s$-wave superconductor, $1/T_1T$ develops a Hebel–Slichter–like peak around $T_{\rm{BKT}}$, originating from the emergence of coherence peaks in the density of states, while no peak appears at the pair formation scale $T_{\rm{BCS}}$. We further extend our analysis to the $d$-wave superconductor. Our results highlight spin-lattice relaxation rate as a sensitive tool to detect the superconducting BKT transition and open routes to exploring its manifestation in unconventional pairing states.

%In high-temperature superconductors, the competition between phase coherence and pairing defines two distinct regimes. In the pseudogap phase, for 
%$T_{\rm{BKT}}<T<T_{\rm{BCS}}$, preformed Cooper pairs lack long-range phase coherence, while below $T_{\rm{BKT}}$ a superconducting state emerges %with well-established phase order. Although the Berezinskii–Kosterlitz–Thouless (BKT) framework captures the nature of this transition, its experimental verification remains challenging because of the absence of sharp thermodynamic signatures. To clarify this issue, we perform classical Monte Carlo simulations of the XY model to probe thermal phase fluctuations in both s-wave and d-wave superconductors. Our results indicate that in the vicinity of 
%$T_{\rm{BKT}}$, s-wave pairing is accompanied by a pronounced peak in the NMR relaxation rate, coinciding with an enhancement of the zero-frequency density of states (DOS). Conversely, while d-wave systems also exhibit an increased zero-frequency DOS above $T_{\rm{BKT}}$, the characteristic NMR peak is absent. We attribute these differences to the distinct low-energy DOS structures dictated by the pairing symmetry. These findings provide concrete numerical signatures to assist in the experimental detection of BKT transitions and in identifying the order parameter symmetry in unconventional superconductors.

\end{abstract}
\maketitle

%\section{Introduction}
Two-dimensional (2D) superconductors provide a fertile ground for studying strong phase fluctuations, where superconductivity can survive in the absence of true long-range order \cite{RevModPhys.78.17,Saito2016,qiu2021recent,keimer2015quantum}. According to the Mermin–Wagner theorem, continuous symmetries cannot be spontaneously broken at finite temperature in two dimensions, ruling out conventional long-range phase coherence \cite{PhysRev.158.383,PhysRevLett.17.1133,Coleman1973}. Nevertheless, thin superconducting films and van der Waals heterostructures exhibit clear superconducting behavior, which is now understood within the framework of the Berezinskii–Kosterlitz–Thouless (BKT) transition \cite{J_M_Kosterlitz_1974,1972JETP_34..610B,J_M_Kosterlitz_1973}.

\begin{figure} % [H] 强制当前位置
    \centering
    \includegraphics[width=3.5in]{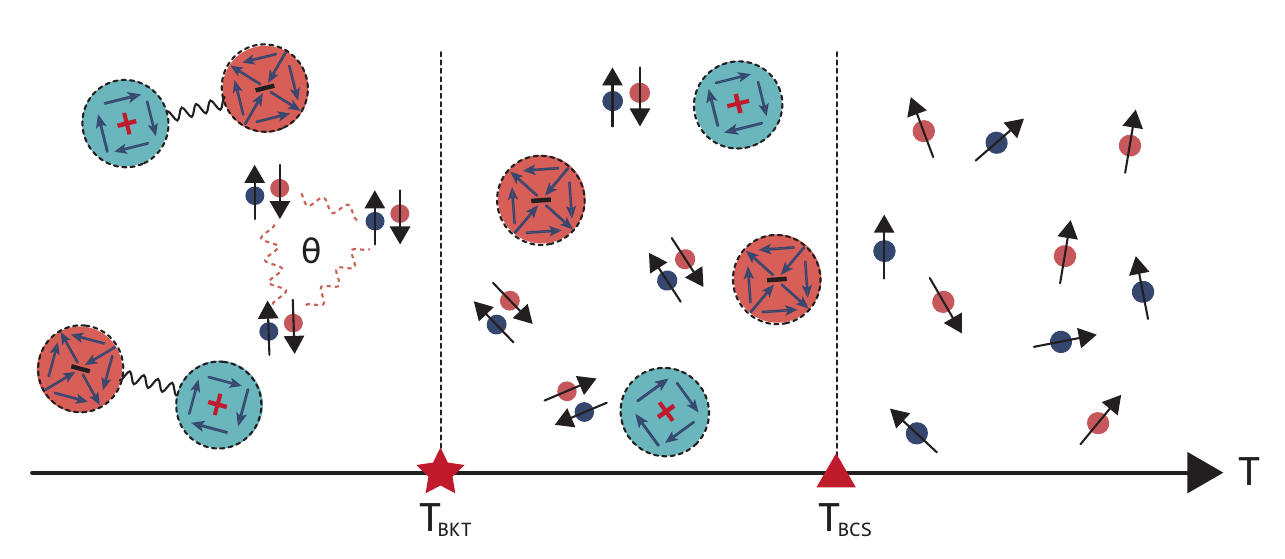} % 单栏宽度
    \caption{
Schematic diagram of the characteristic temperatures \( T_{\mathrm{BKT}} \) and \( T_{\rm{BCS}} \) for the superconducting transition in a phase-fluctuation-dominated unconventional superconductor. 
%For \( T_{\mathrm{BKT}} < T < T_{{BCS}} \), electrons form Cooper pairs, yet strong phase fluctuations prevent global phase coherence and lead to the deconfinement of vortex-antivortex pairs, which can be thermally excited as free vortices. For \( T < T_{{BKT}} \), vortices and antivortices become bound into pairs, and long-range phase coherence is established among the Cooper pairs.
        \label{fig0}
    }
\end{figure}

As schematically illustrated in Fig.~\ref{fig0}, the superconducting transition in two dimensions is characterized by two distinct temperature scales: $T_{\rm{BCS}}$ and $T_{\rm{BKT}}$. Below $T_{\rm{BCS}}$, Cooper pairs begin to form, but strong phase fluctuations prevent the establishment of long-range order. True phase coherence is achieved only at the lower temperature $T_{\rm{BKT}}$, where a collective phase condensation occurs and the system undergoes the BKT transition. This transition is topological in nature, governed by the binding and unbinding of vortex–antivortex pairs. For $T<T_{\rm{BKT}}$, vortices remain bound, leading to quasi-long-range phase coherence and vanishing resistance. In contrast, for $T>T_{\rm{BKT}}$, vortex pairs unbind, proliferating as free vortices that destroy phase coherence and give rise to a finite resistance, even though Cooper pairs persist \cite{Ryzhov:2017,Emery1995,PhysRevB.58.14572,PhysRevB.66.140510,PhysRevB.74.134510,PhysRevLett.99.247001,PhysRevB.82.052503,Li_2011,Li_2011b,li2021superconductor,PhysRevB.107.224502}.

%As schematically illustrated in Fig.\ref{fig1s}, the 2D superconducting transition consists of two important temperatures: $T_{\rm{BKT}}$ and $T_{\rm{BCS}}$. Below $T_{\rm{BCS}}$, Cooper pairs start to form, while the strong phase fluctuation prevents a long-range order. Then, the phase condensation occurs at the temperature of $T_{\rm{BKT}}$, leading to the phase transition of 2D superconductor. 
%On the other hand, the BKT transition is a topological phase transition driven by the binding and unbinding of vortex–antivortex pairs. Below the transition temperature $T_{\mathrm{BKT}}$, vortices remain bound, ensuring quasi-long-range phase coherence and zero electrical resistance. Above $T_{\mathrm{BKT}}$, unbound free vortices proliferate, destroying phase coherence and leading to a finite resistance even though Cooper pairs may remain intact. 

%Experimentally, the identification of the BKT transition relies on the universal jump of the superfluid stiffness at $T_{\mathrm{BKT}}$, as predicted by Kosterlitz and Thouless. This jump is further extended to power-law current–voltage ($V=I^\alpha$) characteristics from Ohmic to non-Ohmic $\alpha\geq3$ \cite{Halperin}. 
%However, it is difficult to observe this $I$-$V$ jump owing to disorder, inhomogeneity etc. Hence, a natural question is other experimental signatures for the superconducting BKT transition.

Experimentally, the BKT transition is most directly identified through the universal jump of the superfluid stiffness at $T_{\mathrm{BKT}}$ \cite{PhysRevLett.40.1727,PhysRevLett.39.1201,PhysRevB.39.2084}. This criterion can be further expressed in terms of the current–voltage characteristics ($V \propto I^\alpha$), where the response evolves from Ohmic behavior to a non-Ohmic power law with exponent $\alpha \geq 3$ \cite{Halperin,Cotón_2011,PhysRevLett.110.196602,ZHAO201359}. In practice, however, the sharp $I$–$V$ jump is often obscured by disorder, inhomogeneity, and finite-size effects etc \cite{PhysRevB.100.064506}. This raises one question of what other experimental signatures can indicate the occurrence of the superconducting BKT transition.

%During the development of Bardeen–Cooper–Schrieffer (BCS) theory, many experimental tools have been applied to verify the BCS pairing hypothesis, including spin-lattice relaxation rate, acoustic attenuation rate, electromagnetic absorption etc.
%Especially, the spin-lattice relaxation rate $1/T_1$ leads to a Hebel-Slichter coherence peak, just below the critical temperature $T_c$, which serves as a crucial experimental validation of the BCS pairing symmetry in conventional superconductors. 
%In this work, we want to explore the spin-lattice relaxation crossing the BKT transition theoretically.

Since the development of Bardeen–Cooper–Schrieffer (BCS) theory, a variety of experimental probes have been employed to test the BCS pairing mechanism, including measurements of the spin-lattice relaxation rate, acoustic attenuation, and electromagnetic absorption \cite{Xiang2022,PhysRev.113.1504,PhysRev.108.1094,tinkham2004introduction}. In particular, the spin-lattice relaxation rate $1/T_1$ exhibits the well-known Hebel–Slichter (HS) coherence peak just below the critical temperature $T_c$, providing a key experimental confirmation of BCS pairing symmetry in conventional superconductors \cite{A_Rigamonti_1998}. Motivated by this, in this work, we investigate theoretically how the spin-lattice relaxation behaves across the BKT transition. We find that in a two-dimensional $s$-wave superconductor, a coherence peak analogous to the HS peak emerges around $T_{\rm{BKT}}$, while no clear feature appears near $T_{\rm{BCS}}$. We further extend our analysis to $d$-wave superconductors.

%In superconductors, the BKT transition manifests through distinct experimental signatures, including the characteristic exponential temperature dependence of resistance, power-law current–voltage ($I$-$V$) characteristics, and the universal stiffness jump. These signatures have been observed in diverse systems such as ultrathin metallic films, oxide interfaces, and van der Waals superconductors. Recent developments in fabricating atomically thin superconductors have renewed interest in the BKT mechanism, providing a platform to explore the interplay between reduced dimensionality, disorder, and strong correlations.

%Despite extensive experimental studies, several open questions remain regarding the precise determination of $T_{\mathrm{BKT}}$, the influence of disorder, and the crossover between amplitude- and phase-driven transitions. Addressing these issues is essential not only for understanding the fundamental physics of low-dimensional superconductivity, but also for designing novel quantum devices where superconducting coherence is controlled at the atomic scale.

%\section{Method and Model}

To capture the essential features of the BKT transition in 2D superconductivity, we consider a two-dimensional Bogoliubov–de Gennes (BdG) Hamiltonian encoding the phase fluctuations governed by an XY model \cite{PhysRevB.66.140510,PhysRevB.82.052503,zhou2024universalscalingbehaviorresistivity}. To simplify our discussion, we start with an s-wave superconductor. Its Hamiltonian is given by:
\begin{eqnarray}
H = &&-t \sum_{\langle i j\rangle, \sigma}\left(c_{i \sigma}^{\dagger} c_{j \sigma}+h.c.\right) -\mu \sum_{ i, \sigma} c_{i \sigma}^{\dagger} c_{i \sigma} \nonumber \\
&& - \sum_{i}\left( \Delta_{i} c_{i \uparrow}^{\dagger} c_{i\downarrow}^{\dagger}+\Delta_{i}^* c_{i\downarrow} c_{i \uparrow} \right),
\label{eq1}
\end{eqnarray}
where $c_{i \sigma}^{\dagger}$ denotes the electron creation operator at site $i$ with spin $\sigma$.
The parameter $t$ represents the effective nearest-neighbor hopping integral and serves as the energy unit ($t=1$) throughout this work. The notation
$\langle i j\rangle$ indicates a summation over nearest-neighbor pairs of lattice sites. 
where $\mu$ is the chemical potential. In this work, to avoid complications from the van Hove singularity on the two-dimensional square lattice, we focus on a system with electron density $n_c=0.15$. At this doping level, the normal state density of states (DOS) near the Fermi level is nearly constant. This choice enables us to systematically investigate the effects of pure phase fluctuations on the physical properties of the system.

The central part of $H$ is the spatial-dependent  SC order parameter $\Delta_{i}$, which has one amplitude part and one phase part as
$\Delta_{i}=|\Delta_i| e^{i \Phi_{i}}$.
Generally speaking, we should expect $T_{\rm{BCS}} \gg T_{\rm{BKT}}$. The spatial fluctuation of amplitude $|\Delta_i|$ is weak around $T_{\rm{BKT}}$. Hence, to capture the essential physics of phase fluctuation to $1/{T_1}$, one can safely take $|\Delta_i|=0.3$ as a constant and focus on the fluctuation of $\Phi_{i}$. We will come back to $|\Delta_i|$ fluctuation later.

\begin{figure} % [H] 强制当前位置
    \centering
    \includegraphics[width=3.7in]{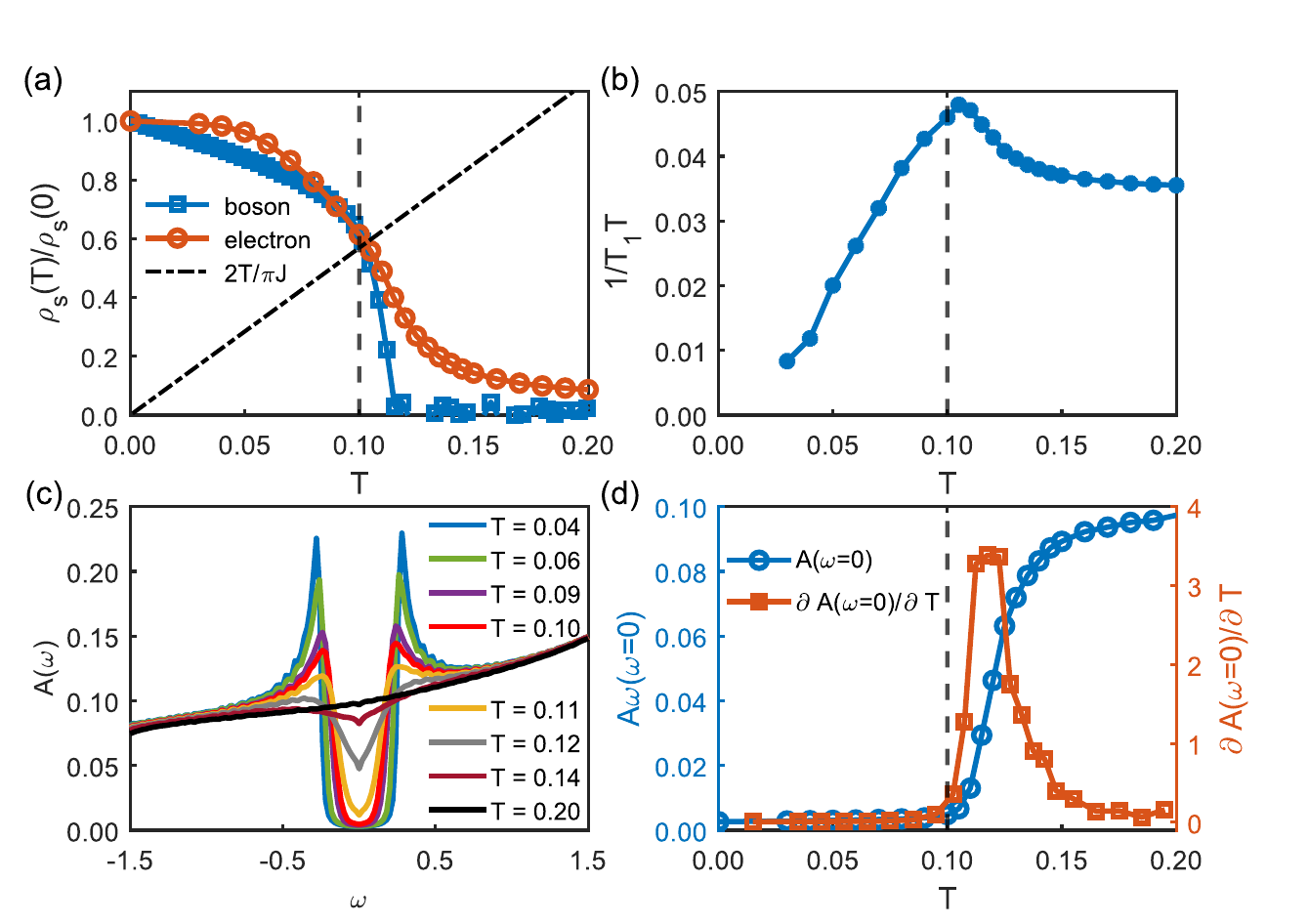} % 单栏宽度
    \caption{
    (a) Temperature dependence of the superfluid density for s-wave pairing. The black dot-dashed curve denotes the universal scaling law $\rho_s(T)/\rho_s(0) \sim 2T/\pi J_{\theta}$, with its intersection with the superfluid curve indicating the BKT transition temperature. Blue squares represent the superfluid density of bosons in the XY model, while red circles show the fermionic superfluid density computed from the BdG Hamiltonian. Although the fermionic and bosonic superfluid densities deviate away from $T_{\rm{BKT}}$, all three curves intersect near $T_{\rm{BKT}}=0.1$. The black dashed line indicates the location of the BKT transition.
    (b) Temperature dependence of the spin–lattice relaxation rate, showing a Hebel–Slichter coherence peak near $T_{\rm{BKT}}$.
    (c) Density of states (DOS) at different temperatures. The U-shaped gap vanishes above the transition temperature.
    (d) Temperature dependence of the zero-frequency DOS $A_{\omega}(\omega=0)$ and its derivative $\frac{\partial A_{\omega}(\omega=0)}{\partial T}$. Below $T_{\rm{BKT}}$
    the U-shaped gap persists and the zero-frequency DOS remains zero. Above the transition temperature, it increases rapidly.
        \label{fig1}
    }
\end{figure}

Then, the phase $\Phi_i$ can be incorporated into an effective XY model and simulated using Monte Carlo methods. The free energy is $F[\Phi]=-J\sum_{ab}\text{cos}(\Phi_a-\Phi_b)$,
where $a$ and $b$ denote the bond between the nearest-neighbor sites for the s-wave pairing state.
The parameter $J$ denotes the interaction strength in the XY model, which yields a well-established BKT transition temperature given by $T_{\rm{BKT}} = 0.893J$ \cite{doi:10.1143/JPSJ.81.113001,app11114931}. Since the superconducting gap $\Delta$ is typically on the order of $\Delta \sim 2 T_{\rm{BCS}}$, we choose $J = 0.112$ such that $T_{\rm{BKT}} = 0.1$, ensuring that it lies well below the mean-field transition temperature $T_{\rm{BCS}}$.
The Monte Carlo simulations are performed on a \(N=72 \times 72\) square lattice. To  further suppress finite-size effects, we employed twisted boundary conditions (see details in the Supplementary Materials).%To further suppress finite-size effects, we employed twisted boundary conditions by introducing one flux quantum in the spin-up and spin-down sectors under the Landau gauge \(\mathbf{A} = (-B y, 0, 0)\) \cite{PhysRevB.65.115104,PhysRevLett.117.097002}. To preserve time-reversal symmetry, fluxes of opposite signs were applied to the spin-up and spin-down components. 

The BKT transition is topological rather than thermodynamic in nature; hence, its transition point cannot be characterized by singularities in specific heat. In Fig.~\ref{fig1}(a), we show the temperature dependence of the superfluid density.
Blue squares represent the superfluid density of bosons in the XY model \cite{Canova_2014}, while
red circles show the fermionic superfluid density \cite{PhysRevLett.68.2830,PhysRevB.47.7995} computed from the BdG Hamiltonian.
The BKT transition temperature is determined by the intersection of the superfluid density curve and the universal line described by the scaling relation \(\rho_s(T)/\rho_s(0) \sim 2T / \pi J\), denoted by the black dashed dot line in Fig.~\ref{fig1} (a). Due to the discrete energy spectrum inherent to a finite fermionic lattice, the numerically computed superfluid density for fermions exhibits deviations from that of the effective bosonic model. Nevertheless, the intersection point with the scaling line remains consistent. Since the phase fluctuations in our model are governed by an XY Hamiltonian, the bosonic phase field \(\{\Phi_i\}\) directly controls vortex formation, pairing, and decoupling processes. Therefore, throughout this work, we use the bosonic superfluid density to locate the BKT transition.

For each configuration of the phase field ${ \Phi }$, the order parameter $\Delta_i$ varies from site to site, breaking the translational invariance of the system. Under such conditions, the spin-lattice relaxation rate is related to magnetic susceptibility \( \chi_{zz} \) via \cite{Xiang2022,slichter2013principles}:
\begin{equation}
    \frac{1}{T_1 T} = \frac{1}{N}\sum_i \frac{2 k_B}{\gamma_e^2 h^3} \sum_{j, 1} F_{j, i} F_{1, i} \lim_{\omega \to 0} \frac{\operatorname{Im} \chi_{zz}(j, 1, \omega)}{\omega},
\label{eq7}
\end{equation}
where \( j \) and \( l \) denote the coordinates of site \( i \) or its neighboring lattice points. Here, \( F_{j,i} \) represents the structure factor of the hyperfine interaction \cite{slichter2013principles}. To reduce computational cost, we include only the dominant contribution from the term with \( j = l = i \).
In the limit \( \omega \to 0 \), the magnetic susceptibility \( \operatorname{Im} \chi_{zz}(j, l, \omega) \) can be expressed using the electron Green's function as:
$\lim_{\omega \to 0} \frac{\operatorname{Im} \chi_{zz}(j, l, \omega)}{\omega} = -\frac{\gamma_e^2 \hbar^2}{2\pi} \int_{-\infty}^{\infty} \mathrm{d}\varepsilon  A(j, l, \varepsilon) \frac{\partial f(\varepsilon)}{\partial \varepsilon}.$
Here $A(j, j', \varepsilon) = \left[ \operatorname{Im} G_{11}(j, j'; \varepsilon) \right]^2 + \left[ \operatorname{Im} G_{12}(j, j'; \varepsilon) \right]^2$, where the Green's functions $G_{11}$ and $G_{12}$ represent the normal and anomalous Green's function, respectively.
Thus $1/T_1T$ can be further rewritten as:
\begin{equation}
\frac{1}{T_1 T} = -\frac{1}{N}\sum_i \frac{k_B}{\pi \hbar} \int \mathrm{d}\varepsilon  \frac{\partial f(\varepsilon)}{\partial \varepsilon} \sum_{j, l} F_{j, i} F_{l, i} A(j, l; \varepsilon).
\label{eq10}
\end{equation}

Fig.~\ref{fig1} (b) shows the calculated spin-lattice relaxation rate $1/T_1T$. At high temperatures, above roughly $2T_{\rm{BKT}}$, $1/T_1T$ remains nearly constant. As the temperature approaches $T_{\rm{BKT}}$, a gradual increase develops, culminating in a pronounced peak near the BKT transition. This peak closely resembles the HS coherence peak observed in conventional superconductors \cite{MACLAUGHLIN19761,tinkham2004introduction,A_Rigamonti_1998,schrieffer2018theory,PhysRev.107.901,PhysRev.113.1504,PhysRev.116.79}. 
Since this system remains a disordered system, the peak is broader than BCS theory.
At lower temperatures, $1/T_1T$ decreases gradually and vanishes as the system enters the phase-coherent superconducting state.

%The result of the spin-lattice relaxation $1/T_1T$ is plotted in Fig.~\ref{fig1} (b).
%We can find that $1/T_1T$ is almost a constant at high temperatures above $2T_{\rm{BKT}}$.
%Then, $1/T_1T$ starts to slightly increase as the temperature approaches $T_{\rm{BKT}}$.
%This increase evolves into a peak near the BKT transition. This feature is consistent with the Hebel–Slichter coherence peak in conventional superconductors. When $T$ is even lower, $1/T_1T$ tends to zero.
%This peak appears at $T \gtrsim T_{\rm{BKT}}$. 

%Physically, the origin of the HS peak comes from the square-root DOS singularity of Bogoliubov quasiparticles \cite{schrieffer2018theory}. This large DOS enhances the spin-lattice relaxation around $T_c$ in BCS theory. On the other hand, at low enough temperatures, few quasi-particles are excited so that the relaxation rate goes to zero as $T -> 0$.

Physically, the origin of the HS peak can be traced to the singular behavior of the quasiparticle DOS in BCS theory. When the superconducting gap opens below $T_c$, the Bogoliubov quasiparticle spectrum produces a square-root divergence at the gap edge \cite{schrieffer2018theory}. This coherence-enhanced DOS strongly amplifies the available scattering channels for nuclear spins, thereby increasing the spin-lattice relaxation rate just below the transition \cite{slichter2013principles}. The enhancement is further reinforced by the BCS coherence factors, which govern the quasiparticle matrix elements entering the relaxation process \cite{schrieffer2018theory}. Together, these effects generate the characteristic HS peak that serves as a hallmark of conventional $s$-wave superconductivity \cite{schrieffer2018theory}. At lower temperatures, however, the number of thermally excited quasiparticles decreases rapidly as the superconducting gap fully develops. Since the relaxation rate is proportional to the quasiparticle population, $1/T_1T$ drops and vanishes in the zero-temperature limit, reflecting the absence of low-energy excitations in a fully gapped $s$-wave state.

It is therefore natural to examine how the quasiparticle DOS evolves with temperature in order to clarify the origin of our observation. The DOS at representative temperatures is shown in Fig.~\ref{fig1}(c), which is calculated by $\text{DOS}(\omega)=-\frac{1}{N\pi}\text{Im}\sum_iG_{11}(i,i,\omega)$.
At $T=0.20$, the spectrum remains nearly flat, closely resembling the normal-state DOS of a two-dimensional square lattice at filling $n_c = 0.15$. This indicates that, although Cooper pairs are already formed, strong phase fluctuations prevent the development of a superconducting gap, leaving the DOS essentially unchanged from its normal-state form. Upon lowering the temperature, a dip–hump structure gradually appears around zero energy, and the DOS acquires a V-shaped profile with weak shoulders near $E \approx 0.24$. Once the system passes through $T=0.10 = T_{\rm{BKT}}$, a clear U-shaped gap emerges, as highlighted by the red curve in Fig.~\ref{fig1}(c). At still lower temperatures, the U-shaped gap deepens, and sharp coherence peaks develop at the gap edge, signaling the onset of robust phase coherence. As discussed above, the HS peak originates from the DOS singularity. In Fig.~\ref{fig1}(c), we see that coherence peaks begin to develop near $T_{\rm{BKT}}$. This suggests that the enhancement in the spin-lattice relaxation rate should also emerge around this temperature, in line with our numerical observations.

It is also useful to analyze the temperature evolution of the zero-frequency DOS \( A_{\omega}(\omega=0) \) and  its derivative \( \frac{\partial A_{\omega}(\omega=0)}{\partial T} \), which hallmarks the transition from V-shape to U-shape DOS.
These results are plotted in Fig.~\ref{fig1}. 
Around \( T \sim 2T_{\rm{BKT}}\), the line shape of the DOS closely resembles that of a conventional metal. As the temperature decreases, the system enters a pseudogap state, where \( A_{\omega}(\omega=0) \) decreases monotonically with decreasing temperature. It eventually vanishes below \( T_{\rm{BKT}} \).
On the other hand, \( \frac{\partial A_{\omega}(\omega=0)}{\partial T} \) shows a sharp peak T=0.118.
This temperature coincides with the peak temperature in $1/T_1T$. 
%Previous theoretical studies have demonstrated that the peak in the temperature derivative of the spectral function at zero frequency, $\frac{\partial A(\omega=0)}{\partial T}$, is more closely associated with a thermodynamic phase transition than with the BKT transition itself \cite{Qin2025}. 

%Our results closely resemble the findings from prior work on overdoped Bi2212, which first identified the singular coincidence between $T_{\rm{BKT}}$ and the peak in the temperature derivative of \( A_{\omega}(\omega=0) \) for the BKT transition \cite{Chen2022}.

\begin{figure} % [H] 强制当前位置
    \centering
    \includegraphics[width=3.1in]{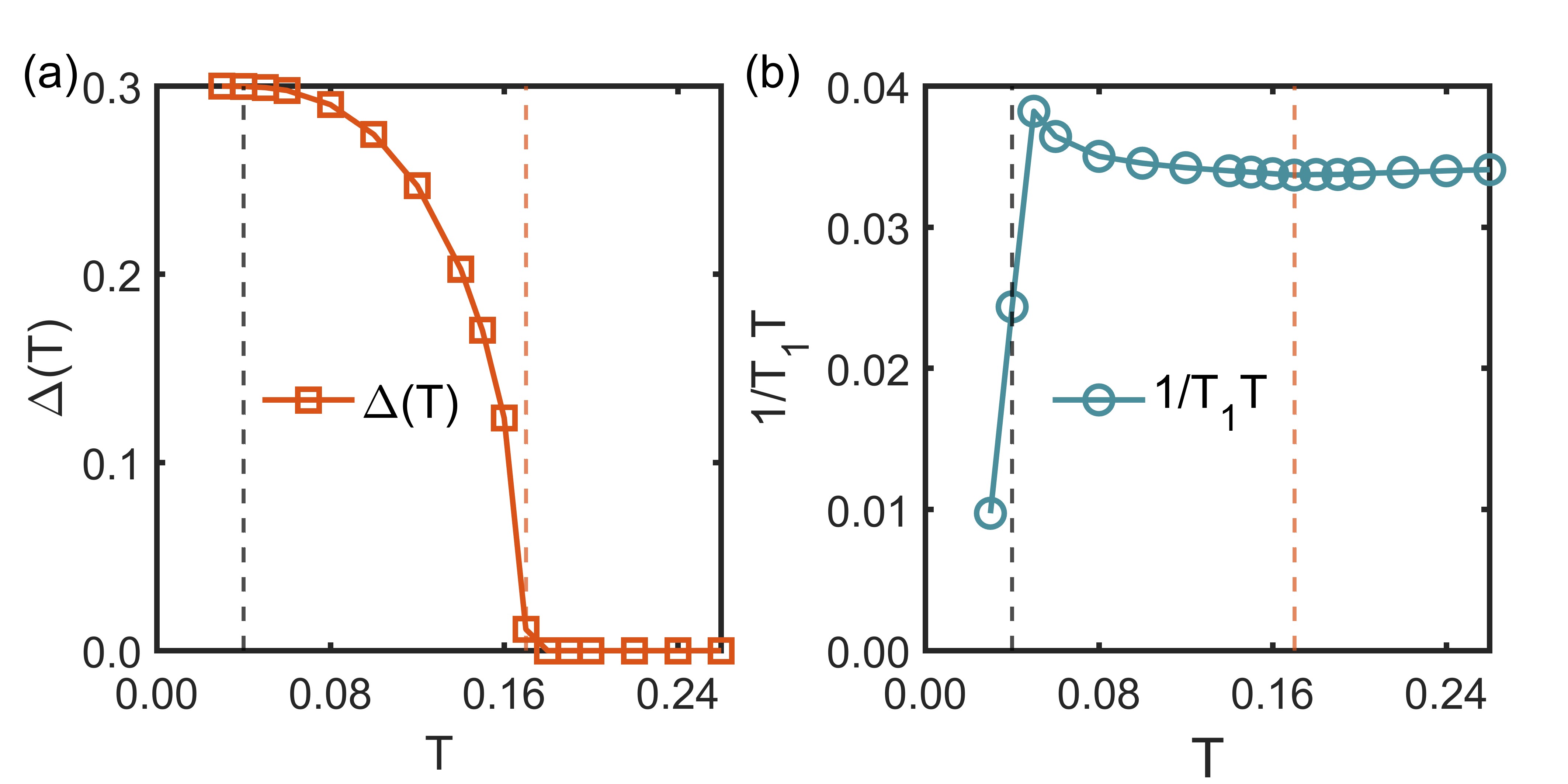} % 单栏宽度
    \caption{
 The temperature-dependent gap magnitude $|\Delta|(T)$ (a) and the corresponding spin-lattice relaxation rate $1/T_1T$ (b) for a system with a $|\Delta|(T)$ determined by self-consistent gap equation. 
$|\Delta|(T)$ is denoted red squares, and $1/T_1T$ by green circles. The BKT transition temperature $T_{\rm{BKT}}$ is indicated by the black dashed line, while the mean-field temperature 
$T_{\rm{BCS}}$, corresponding to gap closure, is marked by the red dashed line.
        \label{fig2}
    }
\end{figure}

Hence, we can conclude that there is an HS-like peak in $1/T_1T$ during the 2D superconducting BKT transition. This also originates from the DOS singularity occurring around $T_{\rm{BKT}}$.
Furthermore, we need to check the $1/T_1T$ feature around $T_{\rm{BCS}}$ to confirm our results. Then, we can include the gap amplitude fluctuation through a BCS self-consistent gap equation \cite{gap_function}.
To focus on the $T_{\rm{BCS}}$ region and ensure that the BKT transition temperature lies well below $T_{\rm{BCS}}$, we adopt a reduced XY-model interaction corresponding to $T_{\rm{BKT}}= 0.04$.
The gap function $|\Delta|(T)$ with $T_{\rm{BCS}}=0.17$ is plotted in Fig.~\ref{fig2}(a). Below $T_{\rm{BCS}}$, the $|\Delta|(T)$ becomes finite and saturates at low temperature following a standard BCS feature.
Correspondingly, $1/T_1T$ is also calculated within the same framework, shown in Fig.~\ref{fig2}(b). We can find that the $1/T_1T$ shows no peak or any distinct feature crossing $T_{\rm{BCS}}$. %which is consistent with our previous observation that DOS remains a normal state around this temperature.
This confirms that the coherence peak observed in our model is exclusively a signature of the BKT transition, driven by the establishment of long-range phase coherence, and is not associated with the formation of Cooper pairs at $T_{\rm{BCS}}$. 
In contrast to the featureless behavior of $1/T_1T$, the Knight shift exhibits a distinct kink at $T_{\rm{BCS}}$ for both $s$- and $d$-wave symmetries, followed by a linear decrease confined to the region immediately below $T_{\rm{BCS}}$ \cite{supp_knight}.

\textit{$d$-wave}
We can also extend our study to $d$-wave superconductors and explore the $1/T_1T$ response crossing $T_{\rm{BKT}}$.
It is well-established that no HS peaks are observed in cuprates \cite{A_Rigamonti_1998}. 
The pairing term for $d$-wave SC of the Hamiltonian Eq.~\eqref{eq1} is expressed as:
\begin{equation}
\begin{gathered}
H_\Delta=  - \sum_{i \delta}\Delta_{i \delta} \left(c_{i \uparrow}^{\dagger} c_{i+\delta \downarrow}^{\dagger}-c_{i \downarrow}^{\dagger} c_{i+\delta \uparrow}^{\dagger}\right)+h.c.,
\end{gathered}
\end{equation}
Different from the s-wave pairing, we should include the $d$-wave pairing phase into $\Phi$. Thus, $\Phi$ denotes the phase of bond ends instead.
The pairing is axis-dependent, where 
$\Phi_{i, \delta=\pm x}=\left(\varphi_i+\varphi_{i+\delta}\right) / 2$, and $\Phi_{i, \delta=\pm y}=\left(\varphi_i+\varphi_{i+\delta}\right) / 2+\pi$. 

%Considering the axis-dependent pairing feature, $\Phi_{i \delta}$ could be explicitly defined as:
%\begin{equation}
%\Phi_{i \delta}=\left\{\begin{array}{lll}
%\left(\varphi_i+\varphi_{i+\delta}\right) / 2 & \text { for } %\delta \text { in } x \text { direction } \\
%\left(\varphi_i+\varphi_{i+\delta}\right) / 2+\pi & \text { for %} \delta \text { in } y \text { direction } .
%\end{array}\right.
%\end{equation}

\begin{figure} % [H] 强制当前位置
    \centering
    \includegraphics[width=3.7in]{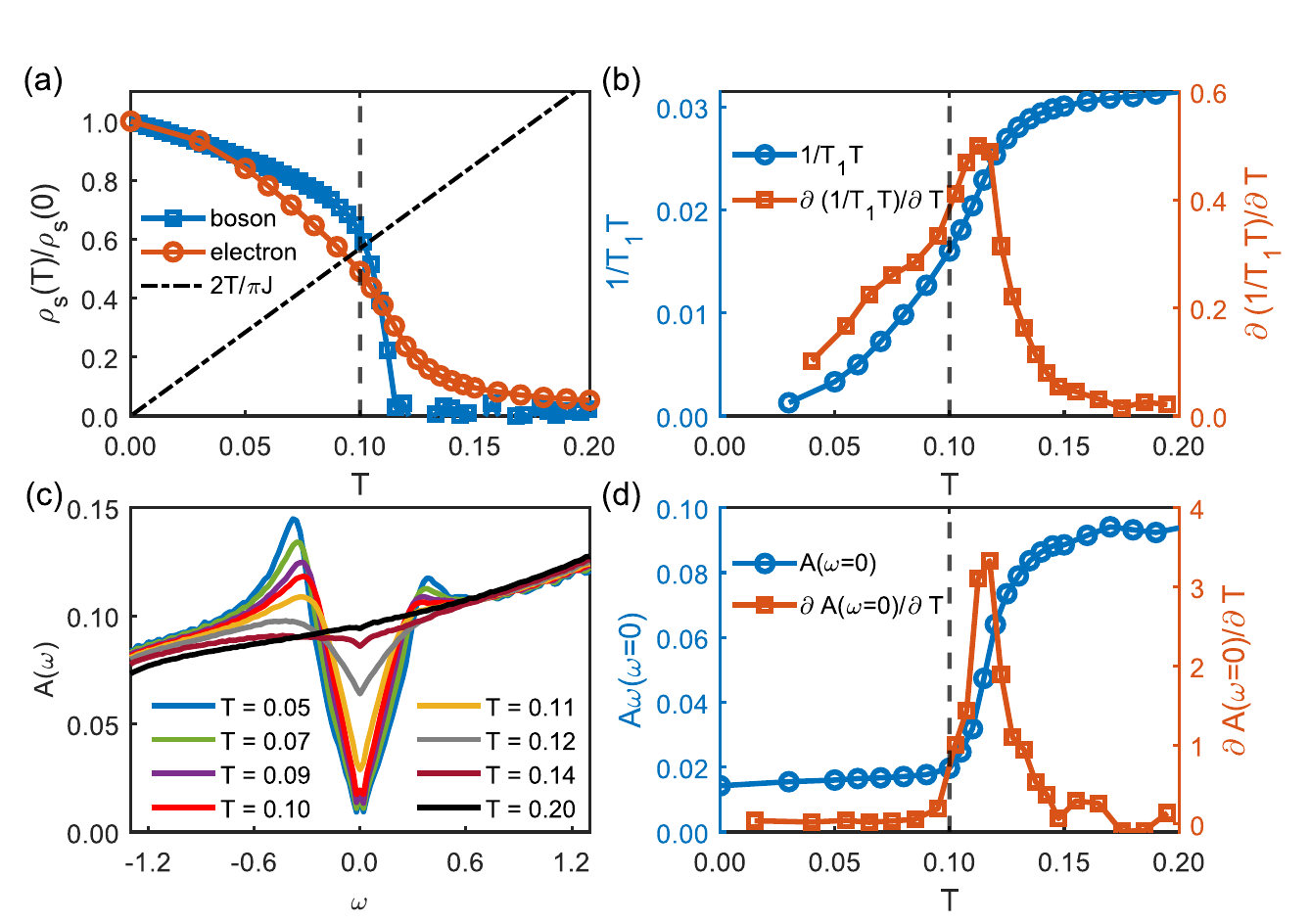} % 单栏宽度
    \caption{
(a) Temperature dependence of the superfluid density for d-wave pairing.  All three curves intersect near $T_{\rm{BKT}}=0.1$. The black dashed line indicates the location of the BKT transition.
(b)  Temperature dependence of the spin–lattice relaxation rate $1/T_1T$ and its derivative with temperature $\frac{\partial (1/T_1T)}{\partial T}$, showing no Hebel–Slichter coherence peak near $T_{\rm{BKT}}$.
(c) DOS at different temperatures.
(d) Temperature dependence of the zero-frequency DOS and its derivative. Below $T_{\rm{BKT}}$ the V-shaped gap persists and the zero-frequency DOS remains nearly zero. Above the transition temperature, it increases rapidly.
\label{fig3}
    }
\end{figure}

We follow the same strategy in the above discussion by fixing the pairing amplitude.
In Fig.~\ref{fig3} (a), we compute the temperature dependence of the superfluid density for a $d$-wave superconductor and determine the BKT transition temperature at \( T_{\rm{BKT}} \sim 0.1 \). 
The behavior of spin-lattice relaxation rate (blue circle) with varying temperature is demonstrated in Fig.~\ref{fig3} (b).
Unlike the s-wave case, although \( 1/T_1T \) remains nearly constant around \( T \sim 2T_{\rm{BKT}} \), however, it drops rapidly near \(T_{\rm{BKT}} \) and exhibits no peak.
Instead, a clear signature of the transition is observed in the temperature derivative, $\partial (1/T_1T) / \partial T$, which displays a sharp peak at around $T_{\rm{BKT}}$ at $T \approx 0.113$.

To further understand the discrepancy of the spin-lattice relaxation rate between the cases of $s$-wave and $d$-wave pairing, we plot the DOS at several temperatures in Fig.~\ref{fig3} (c). In contrast to the s-wave case, the DOS exhibits a stable V-shaped gap. For \( T < T_{\rm{BKT}} \), coherence peaks emerge and their amplitude increases with decreasing temperature. 
The absence of HS peak, which instead emerges in its derivation with temperature, can be understood by analyzing three distinct temperature regimes.
Below \(T_{\rm{BKT}}\), there exists only a smaller spectral contribution near the coherence peaks, and the zero-frequency DOS remains almost unchanged; thus, the slight increase in \(1/T_1T\) with increasing $T$ is mainly due to thermal effects. Above \(T_{\rm{BKT}}\), the coherence peaks vanish completely. The increase in the in-gap spectral weight leads to a gradual increase in \(1/T_1T\). Once the pseudogap is fully suppressed, \(1/T_1T\) approaches a constant value, consistent with the behavior of a conventional metal at high temperatures. In the high-$T$ regime (\(T > T_{\rm{BKT}}\)), the temperature dependence of \(1/T_1T\) fully correlates with that of \(A_\omega(\omega = 0)\). Crucially, this correspondence extends to their respective temperature derivatives. As shown in Fig.~\ref{fig3} (d), there also exists a peak around $T\sim 0.118$.
The zero-frequency DOS \( A_\omega(\omega = 0) \) and its derivative \( \frac{\partial A_\omega(\omega = 0)}{\partial T} \) reveal a gradual emergence and an increase of in-gap spectral weight within the pseudogap regime for \(  T_{\rm BKT}<T<T_{\rm BCS} \). 
%The discrepancy emerges at low temperature regime $T<T_{\rm BKT}$, where \( A_\omega(\omega = 0) \) remains zero until the BKT transition just like the s-wave pairing situation. 
This theoretical result is strikingly similar to experimental observations in the cuprate superconductor Bi2212\cite{Chen2022}. Although its transition is not of the BKT type, Bi2212 is also characterized by pre-formed pairs above $T_c$ whose coherence is destroyed by strong phase fluctuations. Critically, a peak in \( \frac{\partial A_\omega(\omega = 0)}{\partial T} \) is also observed experimentally at $T_c$ in this material, reinforcing its interpretation as a clear signature of the onset of macroscopic phase coherence.

%\section{Conclusion}

Based on extensive Monte Carlo simulations of phase fluctuations within the XY model framework, we have systematically explored the BKT transition in both $s$-wave and $d$-wave superconductors. The superfluid density, extracted from the current-current correlation function, exhibits universal scaling that permits an accurate determination of the BKT transition temperature, $T_{\rm{BKT}}$.
Importantly, the transition driven by phase fluctuations manifests distinctly different features in the spin-lattice relaxation rate, \(1/T_1T\), depending on the pairing symmetry. In s-wave superconductors, \(1/T_1T\) displays a sharp peak near $T_{\rm{BKT}}$, whereas in d-wave systems it decreases monotonically without a discernible peak. Additionally, the zero-frequency DOS is strongly suppressed below $T_{\rm BKT}$ and rises abruptly above $T_{\rm BKT}$ in both cases, with its derivative \( \frac{\partial A_{\omega}(\omega=0)}{\partial T} \) peaking at $T_{\rm BKT}$. 
These findings provide clear numerical signatures for detecting the BKT transition and underscore a direct connection between phase coherence, spectral properties, and NMR relaxation behavior in two-dimensional superconductors. 
Our results offer not only quantitative benchmarks for experimental identification of the BKT transition but also a theoretical framework for distinguishing pairing symmetries through NMR measurements.
We expect that the spin-lattice relaxation discussed here represents just one example, and that other experimental probes originally developed to test BCS pairing may also be extended to explore BKT physics.

The physics of strong phase fluctuations is not limited to two-dimensional systems; it is a hallmark of many unconventional superconductors, including the cuprates. In these materials, Cooper pairs form at a higher pairing temperature $T_{\rm gap}$, but strong long-range superconductivity is only established at a lower temperature $T_c$, upon the onset of phase coherence. This separation of energy scales suggests that our conclusions should hold more generally, even when the transition is not strictly of the BKT type. Namely, spectroscopic signatures of coherence, such as a peak in the spin-lattice relaxation rate (\(1/T_1T\)) or its temperature derivative, should be tied to phase coherence transition of $T_c$, rather than the opening of pairing gap at $T_{\rm gap}$. This insight provides a new interpretive framework for experimental studies, particularly for NMR measurements in materials with a pre-formed pairing due to strong phase fluctuation. A deeper theoretical investigation of this phenomenon in specific three-dimensional models is a promising direction for future research.

\textit{Acknowledgments}
We would like to thank Yin Zhong for useful discussions. We acknowledge the support by the Ministry of Science and Technology  (Grant No. 2022YFA1403900), the National Natural Science Foundation of China (Grant NSFC-12494594, No. NSFC-12174428, No. 12474146, No. 12347107), the New Cornerstone Investigator Program, the Chinese Academy of Sciences Project for Young Scientists in Basic Research (2022YSBR-048), and the Beijing Natural Science Foundation (No. JR25007).
W. Y. acknowledge support from the China Postdoctoral Science Foundation (Grant No. 2025M773414).

\bibliography{main_bib}

\clearpage
\onecolumngrid
\begin{center}
\textbf{\large Supplemental Material: Spin-Lattice Relaxation in Two-Dimensional Superconducting BKT Transition}
\end{center}

%%%%%%%%%% Prefix a "S" to all equations, figures, tables and reset the counter %%%%%%%%%%
\setcounter{equation}{0}
\setcounter{figure}{0}
\setcounter{table}{0}
\setcounter{page}{1}
\makeatletter
\renewcommand{\theequation}{S\arabic{equation}}
\renewcommand{\thefigure}{S\arabic{figure}}
\renewcommand{\thetable}{S\arabic{table}}
%\renewcommand{\bibnumfmt}[1]{[S#1]}
%\renewcommand{\citenumfont}[1]{S#1}
%%%%%%%%%% Prefix a "S" to all equations, figures, tables and reset the counter %%%%%%%%%%

\onecolumngrid

\subsection{The details of Monte Carlo simulation}

Our numerical approach is a hybrid method that combines a classical treatment of order parameter fluctuations with an exact quantum description of the fermions. The thermal fluctuations of the superconducting phase are modeled using a long-wavelength XY model and sampled via standard Monte Carlo (MC) simulations. For each classical phase configuration generated, we determine the single-particle fermionic spectrum by numerically diagonalizing the corresponding quadratic Hamiltonian. While the repeated diagonalization of the fermionic Hamiltonian constitutes the primary computational bottleneck, this hybrid scheme is highly efficient, enabling the study of large systems up to $N=72\times72$ lattice  sites. To minimize finite-size effects, we implement twisted boundary conditions by threading a magnetic flux through the lattice for both spin sectors. This is accomplished with a Peierls substitution in the Landau gauge, $\mathbf{A} = (-B y, 0, 0)$~\cite{PhysRevB.65.115104,PhysRevLett.117.097002}, which modifies the nearest-neighbor hopping terms such that the hopping in the $\hat{x}$-direction acquires a position-dependent phase factor:
\begin{equation}
    c^{\dagger}(n_x, n_y)c(n_x+1, n_y) \rightarrow c^{\dagger}(n_x, n_y)c(n_x+1, n_y)\exp\left(-i\frac{2\pi n_y}{L_x L_y}\right),
\end{equation}
and the periodic boundary condition in the $\hat{y}$-direction is twisted as
\begin{equation}
    c^{\dagger}(n_x, L_y)c(n_x, 1) \rightarrow c^{\dagger}(n_x, L_y)c(n_x, 1)\exp\left(i\frac{2\pi n_x}{L_x}\right).
\end{equation}
To preserve global time-reversal symmetry, we apply magnetic fluxes of opposite sign to the spin-up and spin-down fermions. The resulting flux through each plaquette is \( \frac{2\pi}{L_x L_y} \), a quantity that vanishes in the thermodynamic limit. This ensures that the twisted boundary conditions serve as a proper finite-size correction without altering the bulk physics of the system at thermodynamic limit.

%The thermal fluctuations of the underlying order parameter are treated classically within a long-wavelength XY model and are sampled using a standard Monte Carlo (MC) method. 
%For each classical phase configuration generated via MC, the fermionic single-particle spectrum is obtained by numerically diagonalizing the corresponding quadratic Hamiltonian. 
%This hybrid approach significantly enhances computational efficiency, with the primary bottleneck being the repeated exact diagonalization of the fermionic Hamiltonian. 
%It allows us to access systems as large as $N=72 \times 72$. 
%To mitigate finite-size effects, we implement twisted boundary conditions by threading a magnetic flux quantum through the lattice for both spin sectors. 
%This is achieved via a Peierls substitution in the Landau gauge, $\mathbf{A} = (-B y, 0, 0)$~\cite{PhysRevB.65.115104,PhysRevLett.117.097002}, which modifies the nearest-neighbor hopping terms. 
%The hopping in the $\hat{x}$-direction acquires a phase factor,

\subsection{LDOS of the specific cut across in the Monte Carlo sample}

\begin{figure*}[!t]% [H] 强制当前位置
    \centering
    \includegraphics[width=5.8in]{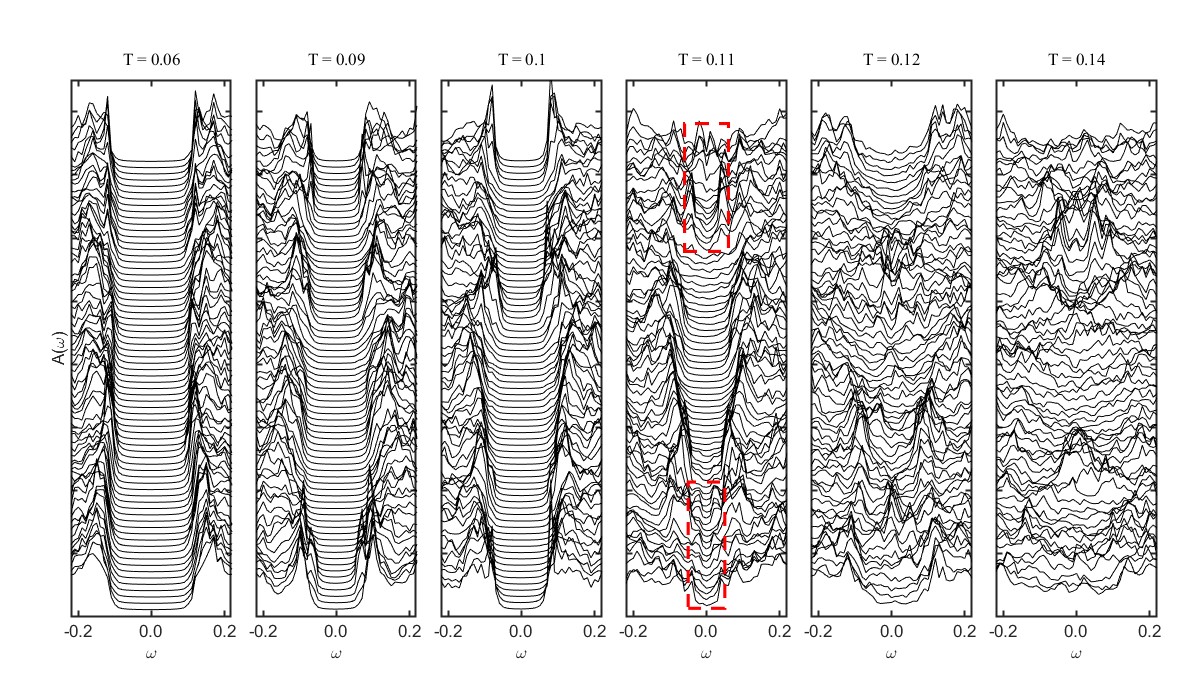} % 单栏宽度
    \caption{
       Waterfall plot of the DOS for s-wave pairing at various temperatures. Each curve represents the local DOS along a horizontal cut across the lattice. As temperature increases, the superconducting gap gradually decreases. Bound states emerge near $T_c$, as highlighted by the red dashed box. Above $T_c$, the gap is smeared out by thermal effects: the U-shaped gap closes and the coherence peak characteristic of s-wave pairing disappears.
        \label{fig1s}
    }
\end{figure*}

To investigate its origin, we plot the waterfall diagram of the local density of states (DOS) along a horizontal cut across the lattice at various temperatures. As illustrated in Fig.~\ref{fig1s}, the absence of translational symmetry leads to spatial variations in the gap magnitude. As temperature increases, the gap size decreases. Near $T_{\rm{BKT}}$, the coherence peaks at the gap edges shift toward zero energy, significantly influencing $1/T_1T$. Below $T_{\rm{BKT}}$, vortex pairs remain confined due to the BKT transition, permitting a finite gap on every site. However, as temperature rises further to $T \gtrsim T_{\rm{BKT}}$, proliferation
 of unbound vortices is expected. The presence of free vortices forces the gap magnitude to vanish at the vortex cores, i.e., $\Delta_i = 0$. This gives rise to finite subgap DOS with small peaks at these sites. Combined with the persistence of coherence peaks at the gap edges, these states further enhance $1/T_1T$, resulting in the observed NMR peak.

\begin{figure*}% [H] 强制当前位置
%    \centering
    \includegraphics[width=5.8in]{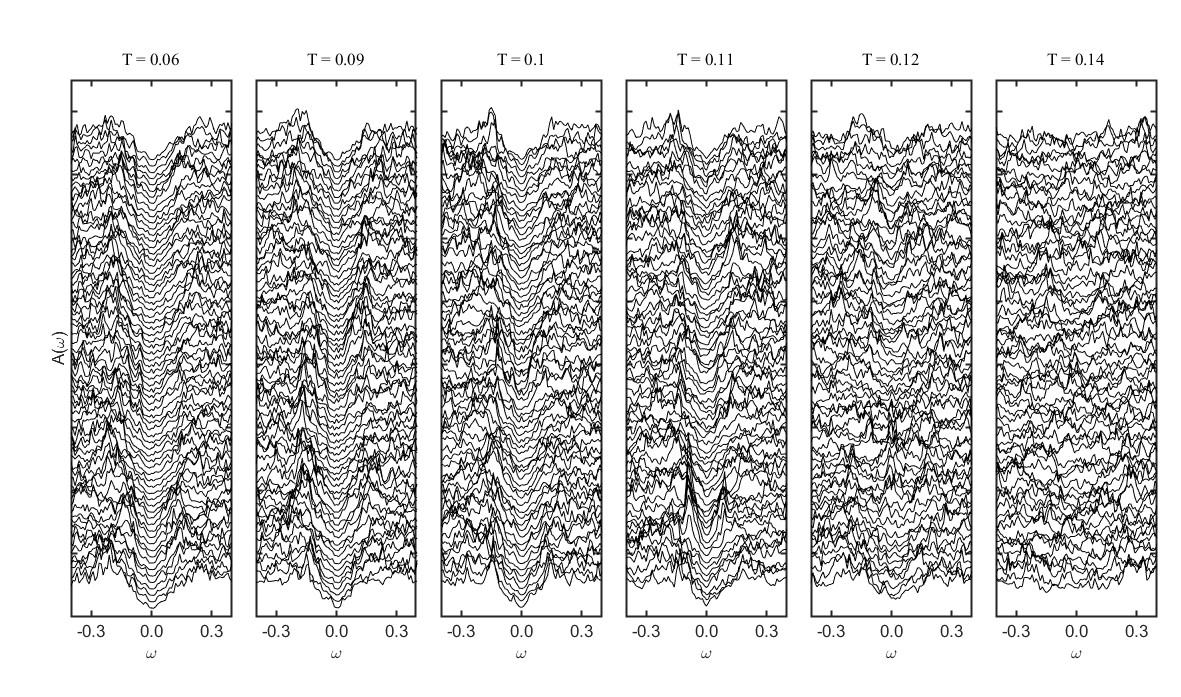} % 单栏宽度
    \caption{
Waterfall plot of the DOS for d-wave pairing at various temperatures. Each curve corresponds to the local DOS along a horizontal cut across the sample. In contrast to s-wave pairing, the DOS near zero energy remains largely unchanged with increasing temperature. However, the coherence peaks gradually diminish below $T_c$ and are strongly suppressed near and above $T_c$.
The combined absence of clear bound states and the suppression of coherence peaks result in no distinctive features in the DOS that would contribute to the spin–lattice relaxation rate, leading to its monotonic increase with temperature.
Above $T_c$ the DOS is smeared by thermal fluctuations.
        \label{fig2s}
    }
\end{figure*}

 Figure \ref{fig2s} presents the waterfall plot of the local DOS for the d-wave superconductor. Combined with Fig.~3(b), these results reveal two distinct regimes in the DOS behavior. First, in the superconducting regime (\(T < T_{\rm{BKT}}\)), the coherence peak amplitude at the gap edges gradually decreases with increasing temperature, where even the zero-temperature coherence peak amplitude is significantly reduced compared to that in the s-wave case. Second, in the pseudogap regime (\(T_{\rm{BKT}} < T < 2T_{\rm{BKT}}\)), the in-gap spectral weight increases steadily until the pseudogap is completely filled.

\subsection{The calculation of the superfluid density}
In the main tex, the superfluid density of bosons in the XY model is calculated by: 
\begin{equation}
\rho_s^{\text{Boson}} = \frac{1}{L^2} \sum_{\langle ab \rangle_x} J \cos (\Phi_a-\Phi_b)  - \frac{\beta}{L^2} \left( \sum_{\langle ab \rangle_x}  J \sin (\Phi_a-\Phi_b) \right)^2,
\end{equation}
where $\beta$ is the inverse temperature. $\langle ...\rangle_x$ denotes the nereast neighbor in x-axis.

On the other hand, we have show that the fermionic superfluid density computed from the BdG Hamiltonian, which is calculated by 
\begin{equation}
    \rho_s^ {\text{Fermion}}=-<K_x>- \Lambda_{xx} (q_x = 0, q_y \to 0, \omega \to 0).
\end{equation}
Here, $K_x$ represent the $x$-component diamagnetic current operator, which can be written as
\begin{equation}
K _ { x } = - \frac { t } { N } \sum _ { \textbf{r} , \sigma } \left( c _ { \textbf{r}  , \sigma } ^ { \dagger } c _ { \textbf{r}+\textbf{e}_x , \sigma } + c _ { \textbf{r}+\textbf{e}_x , \sigma } ^ { \dagger } c _ { \textbf{r} , \sigma } \right).
\end{equation}
The second term $\Lambda_{xx}$ corresponds to the paramagnetic terms of the system, which can be determined by the current-current corelation function
\begin{equation}
\Lambda_{xx}(\mathbf{q}, \omega) = \int_{-\infty}^{\infty} dt e^{i\omega t} \langle [j_x(\mathbf{q},t), j_x(-\mathbf{q},0)] \rangle.
\end{equation}
Here $j^x(\mathbf{q},t)$ is the Fourier component of the current operator along the $x$-direction. In real space the current operator is written as
\begin{equation}
j_x(\mathbf{r}) = it \sum_{\sigma} \left( c_{\mathbf{r},\sigma}^{\dagger} c_{\mathbf{r}+\mathbf{e}_x,\sigma} - c_{\mathbf{r}+\mathbf{e}_x,\sigma}^{\dagger} c_{\mathbf{r},\sigma} \right).
\end{equation}
To calculate the $\Lambda^{xx}(\mathbf{q}, \omega)$, the imaginary-frequency Green function is needed
\begin{align}
\Lambda_{x x} (q, \tau) &= -\langle T_\tau j_x (q, \tau) j_x (-q, 0) \rangle \\
&= -\frac{1}{Z} \mathrm{Tr} \left[ e^{-\beta H} e^{\tau H} j_x (q) e^{-\tau H} j_x (-q) \right] \\
&= -\frac{1}{Z} \sum_{mn} \langle n| e^{-\beta H} e^{\tau H} j_x (q) e^{-\tau H} |m \rangle \langle m| j_x (-q) |n \rangle \\
&= -\frac{1}{Z} \sum_{mn} e^{-\beta E_n} e^{\tau (E_n - E_m)} \langle n| j_x (q) |m \rangle \langle m| j_x (-q) |n \rangle
\end{align}
Thus,
\begin{align}
\Lambda_{xx} (q, \Omega_n) &= \int_0^\beta d\tau e^{i\Omega_n \tau} \Lambda_{xx} (q, \tau) \\
&= -\frac{1}{Z} \sum_{mn} \langle n| j_x (q) |m \rangle \langle m| j_x (-q) |n \rangle e^{-\beta E_n} \int_0^\beta e^{(i\Omega_n + E_n - E_m)\tau} d\tau \\
&= -\frac{1}{Z} \sum_{mn} \langle n| j_x (q) |m \rangle \langle m| j_x (-q) |n \rangle e^{-\beta E_n} \frac{e^{(i\Omega_n + E_n - E_m)\beta} - 1}{i\Omega_n + E_n - E_m} \\
&= -\frac{1}{Z} \sum_{mn} \langle n| j_x (q) |m \rangle \langle m| j_x (-q) |n \rangle e^{-\beta E_n} \frac{e^{(E_n - E_m)\beta} - 1}{i\Omega_n + E_n - E_m} \\
&= \frac{1}{Z} \sum_{mn} \langle n| j_x (q) |m \rangle \langle m| j_x (-q) |n \rangle \frac{e^{-\beta E_n} - e^{-\beta E_m}}{i\Omega_n + E_n - E_m} \\
&= \frac{1}{Z} \sum_{mn} \langle n| j_x (q) |m \rangle \langle m| j_x (-q) |n \rangle \frac{e^{-\beta E_n}}{i\Omega_n + E_n - E_m}
 - \frac{1}{Z} \sum_{mn} \langle n| j_x (q) |m \rangle \langle m| j_x (-q) |n \rangle \frac{e^{-\beta E_m}}{i\Omega_n + E_n - E_m}.
\end{align}

\subsection{Knight shift around BCS transition}
Under an applied magnetic field, the polarization of conduction electron spins generates an additional local field at the nucleus through the hyperfine interaction. The resulting shift in resonance frequency relative to the external field, defined as the Knight shift, is written as \cite{Xiang2022}:
\begin{equation}
K_{z}= \frac{F(0) \, \chi_{zz}(0, 0)}{\gamma_N \gamma_e \hbar^2}.
\end{equation}
Here, the static electron spin susceptibility in the long-wavelength limit can be expressed as:
\begin{equation}
\label{eq:chi_limit}
\lim_{q \to 0} \chi_{zz}(q,0) = -\frac{\gamma_e^2 \hbar^2}{2V} \sum_{k} \frac{\partial f(E_k)}{\partial E_k}.
\end{equation}

\begin{figure} % [H] 强制当前位置
    \centering
    \includegraphics[width=3.7in]{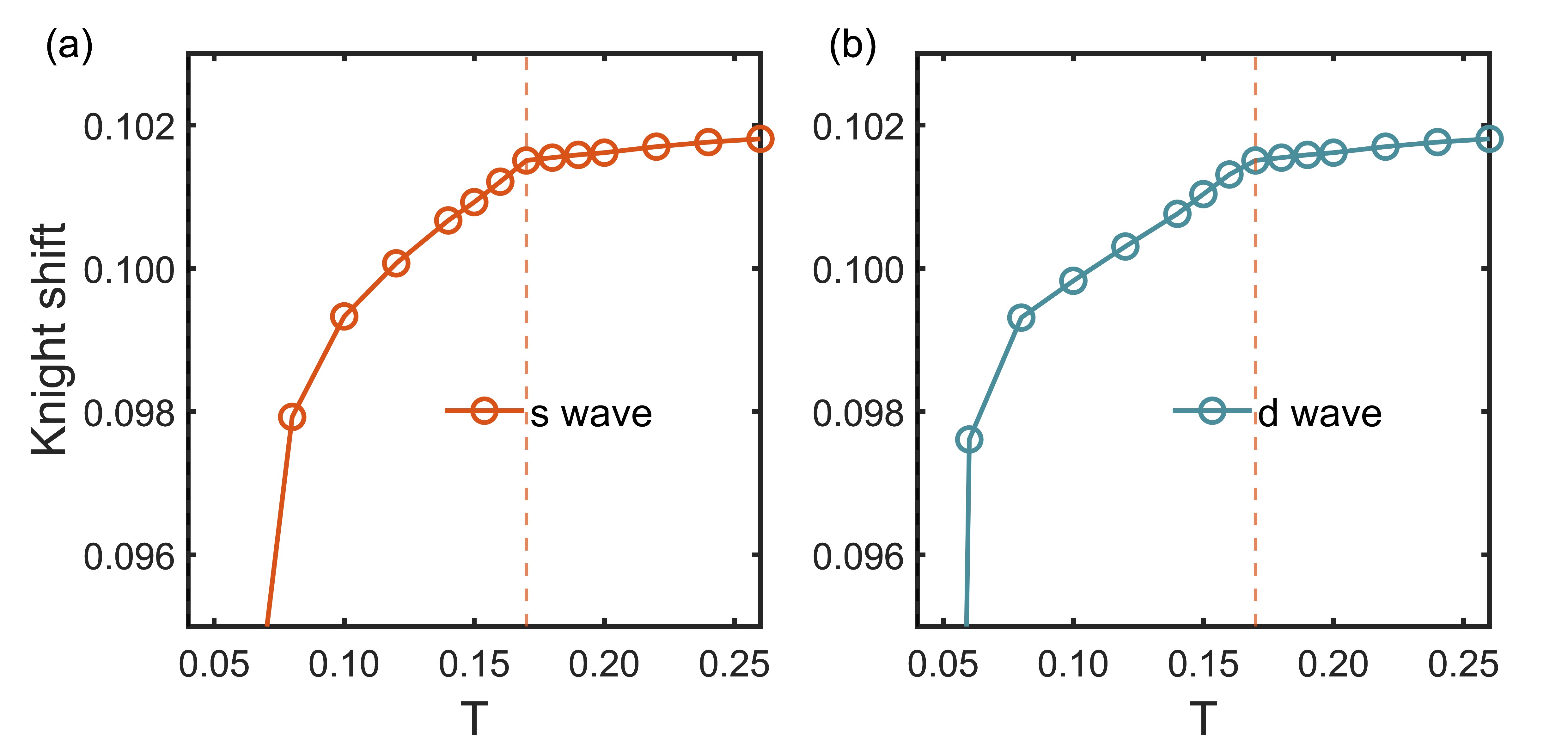} % 单栏宽度
    \caption{
Temperature dependence of the knight shift for s-wave pairing (a) and d-wave pairing (b). The mean-field temperature 
$T_{\rm{BCS}}$, corresponding to gap closure, is marked by the red dashed line.
\label{figS3}
    }
\end{figure}

Here, we examine the Knight shift near $T_{\mathrm{BCS}}$ to search for signals of the BCS transition. 
Following the main text, gap amplitude fluctuations are included through the self-consistent BCS gap equation. 
We employ a reduced XY interaction strength ($T_{\mathrm{BKT}}= 0.04$) to separate the BKT transition well below $T_{\mathrm{BCS}}$. 
As illustrated in Fig.~\ref{figS3}, the Knight shift is nearly constant in the normal state ($T > T_{\mathrm{BCS}}$) due to the constant DOS. 
Upon crossing $T_{\mathrm{BCS}}$, the Knight shift exhibits a linear decrease for both $s$- and $d$-wave cases, resulting in a kink at the transition temperature. 
However, this linear behavior vanishes as the temperature approaches the BKT transition. 
As shown in Fig.~\ref{figS3}, the linear dependence persists down to $T \approx 0.1$ for the $s$-wave symmetry, whereas it extends to $T \approx 0.08$ for the $d$-wave symmetry. 
Below these respective temperatures, the Knight shift decreases more rapidly due to the sharp depletion of the DOS at the Fermi energy.

\end{document}